# Prediction of Component Shifts in Pick and Place Process of Surface Mount Technology Using Support Vector Regression

Shun Cao[a], Irandokht Parviziomran[a], Haeyong Yang[b], Seungbae Park[c], Daehan Won[a]

[a]Department of Systems Science and Industrial Engineering, The state University of New York at Binghamton, Binghamton, NY, USA
[b]Koh Young Technology Inc. Seoul, South Korean
[c]Department of Mechanical Engineering, The state University of New York at Binghamton, Binghamton, NY, USA

**Abstract**

In pick and place (P&P) process of surface mount technology (SMT) the placed component can shift from its ideal (or designed) position on the wet solder paste. The solder paste with some fluid properties could slump and the unbalance between different sides of solder paste can lead to other forces on the components as well. Though the shifts are usually considered to be negligible and can be made up to some extent by the following self-alignment during the process of soldering reflow, it should be attracted attention as its importance for addressing the quality of the printed circuit board (PCB) in SMT. To minimize or control the component shifts, whose relationship with the characteristics of the solder paste (e.g., offset, volume) should be studied initially. In this paper, we design a comprehensive experiment and collect the data from a state-of-the-art SMT assembly line. Then we use support vector regression (SVR) model to predict the component shifts based on different situations of solder paste and placement settings. Also, two kernel functions, linear (SVR-Linear) and radial basis function (SVR-RBF), are employed. The achieved results indicate that the component shift in P&P process is significant, and the SVR model is highly qualified for the forecast of the component shifts. Particularly, the SVR-RBF model outperforms the SVR-Linear model considering the prediction error.

## 1. Introduction

*1.1. Surface Mount Technology*

Surface mount technology (SMT) is well-known as an essential method for electronic component assembly. The main operations in a surface mount assembly (SMA) line are stencil printing process (SPP), pick and place (P&P), and solder reflow (for detail, see Fig. 1). A PCB stencil is aligned on the surface of the boards and solder paste is applied using a squeegee blade to ensure the pads are coated with a controlled amount of solder paste. Then, the components are mounted onto the PCB boards in their respective positions by a P&P machine. Finally, the boards are passed through a reflow oven, in which the flux in the solder paste will evaporate, and the solder paste will be melt into liquid and form solder joints.

Except the three main processes mentioned above, the PCB boards in an SMA line need to be tested respectively by a Solder Paste Inspection (SPI) machine, and one or two Automated Optical Inspection (AOI) machines to evaluate the quality of outcome for each process. Mainly, SPI checks the quality of the solder paste after printing; Pre-AOI, which is located before the reflowing oven, takes charge of testing the placed components after P&P process. Post-AOI is more popular than Pre-AOI. It is laid following the reflow oven, exams the assembled components after reflow soldering.

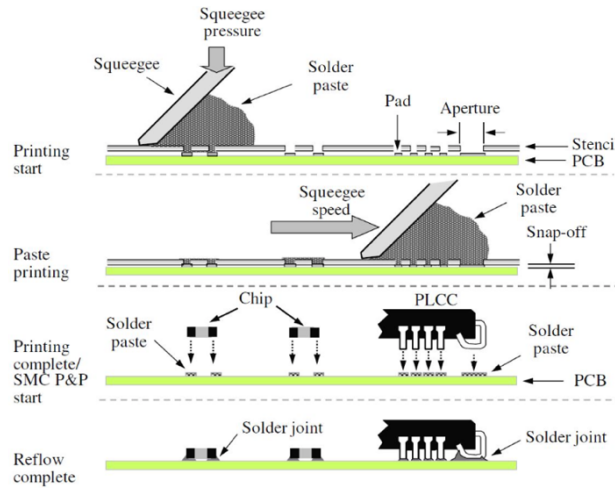

Fig. 1. Main processes of Surface mount technology [1].

*1.2. Component Shift in P&P Process*

The P&P operations are applied on a Surface Mount Device (SMD) and widely used in the electronics industry. When the component is placed on the wet solder paste, which could be considered as a temporary adhesive [2], it is to be held in position by the tackiness of solder paste [3]. However, the position of the placed component might be changed before entering the heating stage as the wet solder paste is a viscous non-Newtonian fluid which slumps more or less during the P&P process [4~6]. Also, if the component is placed poorly on the solder paste or the solder paste is too thin, too thick or too unbalanced etc., other forces will act on the placed component and cause the component shifts on it. Besides, there are many other indirect potential factors that can lead to component shifts, such as machine's vibration, PCB's oblique, conveyor's instability.

In practice, the component shifts in P&P process are often underestimated because its tiny amount of the measured distance and the following reflow soldering is considered as a standard way to make up the shifts and placement error [7]. However, with the desire of decreasing defective rate of assembled products in SMT, and the decreasing size of the electronic components, the component shifts in P&P process are no more ignorable, i.e., the tiny shifts may bring out significant misalignment from the small components. But note that it is difficult to detect the component shifts merely by AOI machine in the SMA line since the difference between designed and tested positions of the component are normlay tangled with the variation from the P&P process, inaccuracy from the inspection equipment, and other hidden environmental situations (e.g., temperature, humidity). In this paper, a comprehensive experiment, which considers a variety of possible situations, even some of them rarely happens in practice, is designed for this study.

*1.3. Support Vector Regression*

Support vector regression (SVR) have attracted significant attention as its importance becomes genuinely recognized for regression purposes [8]. It is extended from support vector machine (SVM) by the introduction of an alternative loss function.

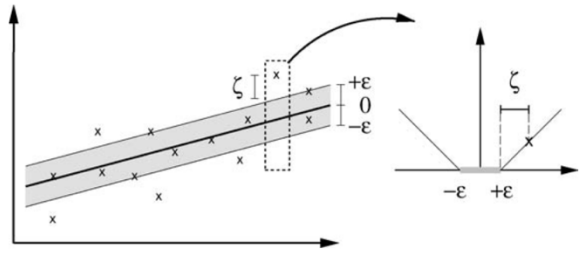

Fig. 2. The soft margin loss setting for linear SVR [8].

The basic idea of SVR prediction model is shown in Fig. 2. Given the training dataset as data $\{(x_1, y_1), ..., (x_l, y_l)\} \in X \times \mathbb{R}$, where $X$ denotes the space of input patterns. SVR's ultimate goal is to find a function $f(x)$ that has at most ε deviation from the actually obtained targets $y_i$ for all the training dataset, and at the same time is as flat as possible [8]. The function of SVR can be written as follows:

$$\begin{aligned}&\min \quad \tfrac{1}{2}\|\omega\|^2 + C\sum_{i=1}^{l}(\xi_i + \xi_i^*) \\ &s.t. \quad y_i - \langle \omega, x_i \rangle - b \leq \varepsilon - \xi_i \\ &\quad\quad\ \langle \omega, x_i \rangle + b - y_i \leq \varepsilon - \xi_i^* \\ &\quad\quad\ \xi_i, \xi_i^* \geq 0 \end{aligned} \quad (1)$$

Where $\langle \cdot, \cdot \rangle$ denotes the dot product, $\|\omega\|^2$ denotes the $L_2$ norm of ω. The slack variables $\xi_i, \xi_i^*$ are to cope with other infeasible constraints of the convex optimization problem for each $i$-th sample. The constant $C > 0$ determines the trade-off between the flatness of $f(x)$ and the amount up to which deviations larger than ε. The flexibility of SVR is in regard to its various kernel functions. Besides, we use two most popular kernel functions as well, one of which is linear kernel function. It is computationally fast and less prone to overfitting. The other one is the radial basis function (RBF), it not only is computational efficient, but also is more suitable to solve the nonlinear problems.

The rest of this paper is structured as follows: the design of experiments, different experimental settings and the detail of acquired datasets are introduced in Sec. 2; the behavior of the component shifts in P&P process is briefly studied in Sec. 3; mathematical process and the prediction results are discussed in Secs. 4; and finally, conclusions are summarized in Sec. 5.

## 2. Experimental Setup and Data Description

### 2.1. Experimental Setup

The experiment was carried out in the field laboratory, which contains a complete state-of-the-art SMT assembly line. The schematic diagram of the SMA line is shown in Fig. 3. One specifically designed PCB board is used in this experiment, on which the pads are coupled with each other corresponding to one electronic component. As represented in Table 1, nine parameters: solder paste offsets $X, Y$, angle, solder paste average volume, the difference of solder paste volume, place pressure, and designed offsets $X, Y$, angle are controlled and adjusted to form 33 different experimental settings according to the theory of design of experiments (DOE). Except that, 6 types of components (shown in Table 2) and 20 replications for each setting are implemented in this experiment as well.

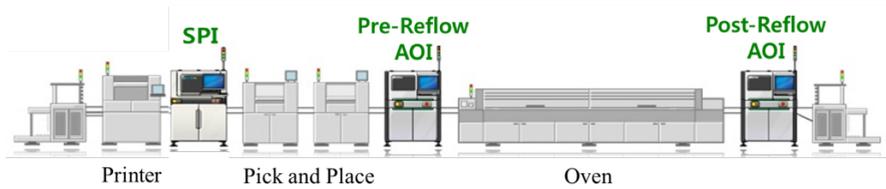

Fig. 3. Surface mount assembly line in the field laboratory.

The experimental flowchart is shown in Fig. 4. Firstly, a specialist of SMT designed the experiment (the solder paste offsets, solder paste volume and placement offsets) based on the D-Optimal designs in DOE. After customizing this piece of PCB board, the solder paste is printed on it according to various solder paste offsets and volumes. SPI machine detects the actual values of solder paste offsets and volumes. Then the PCB board is passed in P&P process, in which the components are placed on the corresponding pair of pads on PCB, based on the settings of part designed offset $X$, part designed offset $Y$, part designed angle and place pressure. Finally, the AOI machine tests the actual component's offsets.

Table 1. Experimental settings of Capacitor Chip 0402.

| Factor | Setting 1 | Setting 2 | Setting 3 | Setting 4 | Setting 5 | … | Setting 33 |
|---|---|---|---|---|---|---|---|
| Solder paste designed offset $X$ ($\mu m$) | 76.84 | 76.84 | 76.84 | 65.92 | 175.00 | … | 141.76 |
| Solder paste designed offset $Y$ ($\mu m$) | 71.12 | 71.12 | 71.12 | 129.56 | 84.00 | … | 220.23 |
| Solder paste designed Angle (°) | -6.92 | 6.92 | 6.92 | -6.92 | 6.92 | … | 6.91 |
| Average volume (%) | 80.00 | 120.00 | 120.00 | 80.00 | 120.00 | … | 120.00 |
| Difference of volume (%) | 0.00 | -40.00 | 0.00 | -40.00 | -40.00 | … | 0.00 |
| Part designed offset $X$ ($\mu m$) | 235.37 | 158.43 | 76.85 | 81.49 | 253.96 | … | 170.73 |
| Part designed offset $Y$ ($\mu m$) | 0.00 | 0.00 | 71.12 | 0.00 | 94.36 | … | 111.81 |
| Part designed Angle (°) | -6.92 | 0.00 | 0.00 | 0.00 | 0.00 | … | 0.00 |
| Place pressure (gram-force) | 150.00 | 0.00 | 150.00 | 150.00 | 0.00 | … | 150.00 |

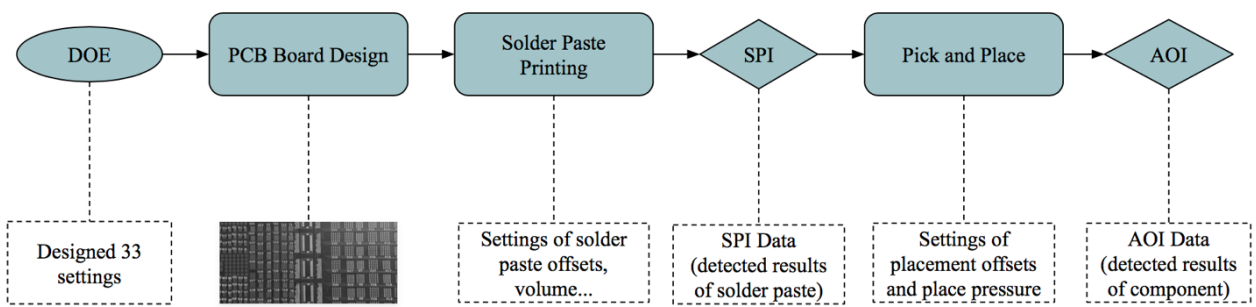

Fig. 4. The experimental flowchart.

Table 2. Components used in the experiment.

| Component Name | Resistor/Capacitor | Dimensions ($\mu m$) | Quantity/board |
|---|---|---|---|
| R01005 | Resistor Chip | 400×200 | 660 |
| R0201 | Resistor Chip | 600×300 | 660 |
| R0402 | Resistor Chip | 1000×500 | 660 |

| C01005 | Capacitor Chip | 400×200 | 660 |
| C0201 | Capacitor Chip | 600×300 | 660 |
| C0402 | Capacitor Chip | 1000×500 | 660 |

## 2.2. Data Description

The amount of the components used in the experiment is 3960 (660×6). The datasets acquired from the SMA line are SPI dataset (size: 7920=3960×2, one component corresponds to two solder paste data) and AOI dataset (size: 3960). All the factors used in this paper are shown in Table 3, in which the offsets are defined as the distances based on the centers: pad center (a pair of pads), solder paste center (a pair of solder paste) and the component center. We illustrate the way of measuring offsets in Fig. 5(a) and Fig. 5(b). Besides, the offsets and volume are designed proportionally based on the size of the component (except angle), therefore, the offsets and volume are transformed to ratios in the prediction model.

Table 3. The brief explanation of the factors.

| Factor | Factor Name | Explanation |
|---|---|---|
| Solder paste offset $X$ ratio | $X_1$ | Solder paste offset in $X$ direction (SPI) / component length |
| Solder paste offset $Y$ ratio | $X_2$ | Solder paste offset in $Y$ direction (SPI) / component width |
| Solder paste angle | $X_3$ | Solder paste rotation (SPI) |
| Average volume ratio | $X_4$ | The average volume of a pair of solder paste / ideal volume |
| The difference of volume ratio | $X_5$ | The difference of volume between a pair of solder paste / ideal volume |
| Part designed offset $X$ ratio | $X_6$ | Component designed offset in $X$ direction (DOE) / component length |
| Part designed offset $Y$ ratio | $X_7$ | Component designed offset in $Y$ direction (DOE) / component width |
| Part designed angle | $X_8$ | Component designed rotation (DOE) |
| Place pressure | $X_9$ | Pressure setting in the P&P machine (DOE) |
| Shift $X$ ratio | $Y_x$ | Part tested offset $X$ ratio (Pre-AOI) – Part designed offset $X$ ratio (DOE) |
| Shift $Y$ ratio | $Y_y$ | Part tested offset $Y$ ratio (Pre-AOI) – Part designed offset $Y$ ratio (DOE) |
| Shift angle | $Y_{ang}$ | Part tested angle (Pre-AOI) – Part designed angle (DOE) |

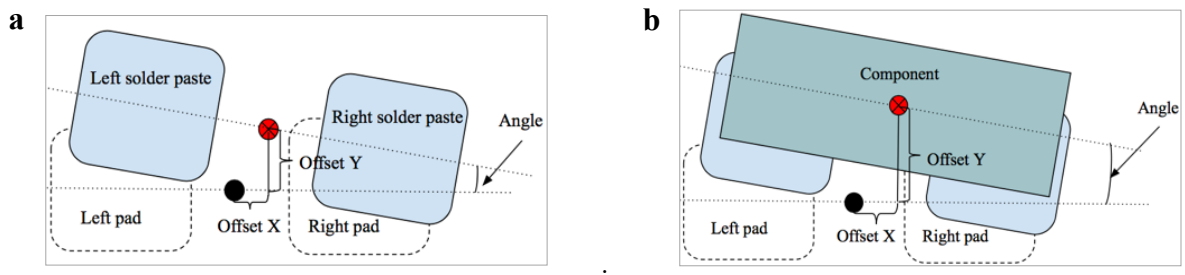

Fig. 5. definition of the offsets: (a) offsets of the solder paste; (b) offsets of the component.

## 3. The Behavior of Component Shifts in P&P Process

It is challenging to investigate the behavior of component shifts in P&P process barely based on the rough data acquired from the SMA line. In this paper, we assume the component shifts are the difference between the designed and tested component positions (for detail, see table 3), though the placement and testing errors are entangled in it. Fig. 6 shows the behavior of the component shifts in P&P process as a case example. In which the black rectangle denotes the designed position of capacitor chip 0402 in setting 1, and the light blue rectangles signify the tested positions of the 20 repeated placements with the same setting. Note that the dimension and position of the rectangles in the figure are depicted according to the actual values in the dataset. The replicated components' positions do not distribute randomly around the designed poison, but tend to distribute in a certain direction. Fig. 7 and Table 4 can show more information of the component shifts in three directions respectively. Apparently, the component shifts in P&P process are significant. A few of shift Ys are even greater than 10% of the component width (500 μm). Moreover, the component shift angles are distributed far from the designed position as well, larger than the designed value 0°.

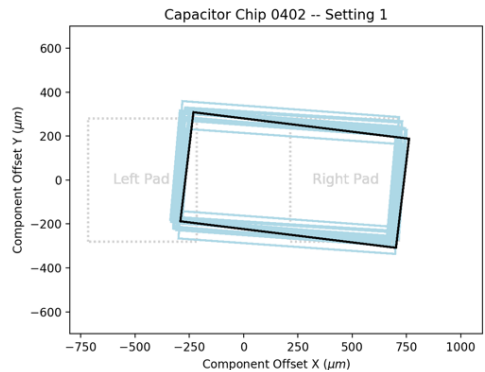

Fig. 6. Behavior of the component positions in P&P process (setting 1 of capacitor chip 0402)

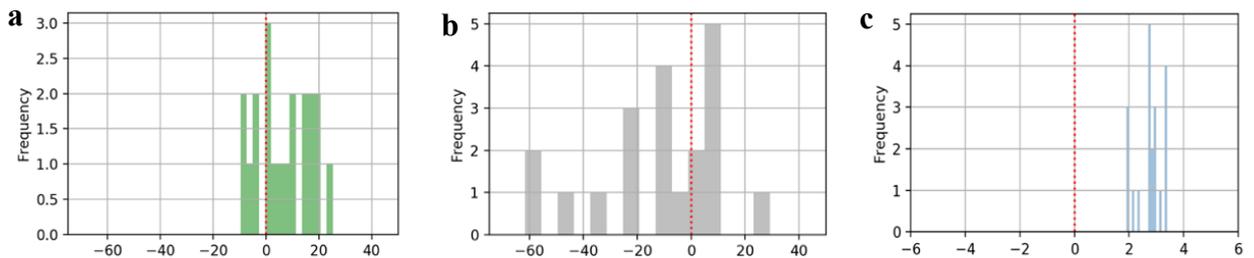

Fig. 7. (a) histogram of component shift $X$ ($\mu m$); (b) histogram of component shift $Y$ ($\mu m$); (c) histogram of component shift Angle (°).

Table 4. Capacitor chip 0402's shifts in P&P process.

| Setting | Shift $X$ ($\mu m$) | | | | Shift $Y$ ($\mu m$) | | | | Shift Angle (°) | | | |
|---|---|---|---|---|---|---|---|---|---|---|---|---|
| | Avg. | Std. | Min. | Max. | Avg. | Std. | Min. | Max. | Avg. | Std. | Min. | Max. |
| 1 | 6.8 | 9.9 | -9.7 | 25.30 | -12.4 | 22.6 | -61.6 | 29.2 | 2.7 | 0.5 | 1.9 | 3.4 |
| 2 | -12.9 | 8.8 | -29.7 | 15.7 | 5.8 | 18.8 | -25.2 | 43.8 | -0.6 | 0.6 | -1.4 | 0.7 |
| 3 | -8.2 | 11.0 | -22.7 | 19.4 | -10.9 | 9.8 | -26.1 | 4.3 | -0.45 | 0.5 | -1.4 | 0.0 |
| … | … | … | … | … | … | … | … | … | … | … | … | … |
| 33 | -7.8 | 9.5 | -22.0 | 13.2 | -2.4 | 21.7 | -46.1 | 48.1 | -0.7 | 0.5 | -1.4 | 0.0 |

## 4. Results

In this section, we estimate component shifts in P&P process based on the 33 settings of DOE using one of the most classical machine learning model: support vector regression (SVR). The selection of kernel function is crucial to SVR's prediction accuracy [9]. In this paper, two most popular kernel functions are utilized with SVR model: linear function (SVR-Linear) and radial basis function (SVR-RBF). In order to evaluate the performance of SVR-Linear and SVR-RBF to recognize whether they are qualified to predict the component shifts, mean absolute error (MAE) and root mean squared error (RMSE) are used to evaluate the prediction error.

Furthermore, since the absence of a huge size of the dataset that can be used to train the prediction model, we apply $k$-fold Cross-Validation ($k$-fold CV) to improve the performance of the prediction model with the limited data. In term of the implementation of $k$-fold CV, the $k$'s value could notably affect the results of the estimation. Therefore, we use model of SVR-RBF and the metric of RMSE to select the proper value of $k$. The results are shown in Fig. 8. It is obvious to see the turning stage of shift $X$ and shift $Y$. However, it hard to discern any turning point in shift angle. Normally, it is recommended that $k = 5$ or $k = 10$ [10]. In this paper, we choose $k = 10$ for the relative lower prediction error, though a little computational efficiency is sacrificed.

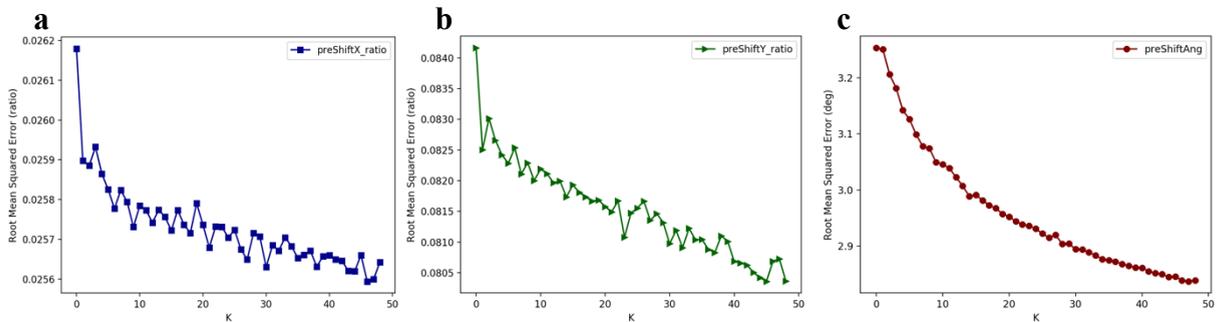

Fig. 8. (a) RMSE across $k$-fold in $x$ direction; (b) RMSE across $k$-fold in $y$ direction; (c) RMSE across $k$-fold in angular direction

Except for the selection of kernel function, SVR's prediction accuracy of is also depended on the tuning of the multiple hyper parameters. The related hyper parameters: $C$ and $\varepsilon$ in SVR-Linear (shown in function 1) and $C$, $\varepsilon$, $\gamma$ SVR-RBF model respectively. $\gamma$ is an important free parameter to decide the variance in SVR-RBF. Notice that, it is time-consuming to select the optimal combination of the three parameters by hand. Fortunately, Grid Search model could help to address this issue, which is simply an exhaustive searching through a manually specified subset of the hyper parameter space of a learning algorithm [11]. In this paper, the selected values of the hyper parameters are $C = 1.0$, $\varepsilon = 0.031$ and $C = 0.13$, $\gamma = 1.0$, $\varepsilon = 0.00097$ for SVR-Linear and SVR-RBF respectively.

Table 5 summarizes the prediction errors (MAE and RMSE) of the component shifts in P&P process using SVR-Linear and SVR-RBF models. The prediction errors are quite small for shift $X$ and shift $Y$, which showed that SVR model is capable of estimating the component shift $X$ and shift $Y$ with high precision levels. It also means that the component shift $X$ and shift $Y$ are explained well by the 9-controlled solder paste and placement parameters in SVR model. Given the condition of solder paste after printing (or the test results of SPI) and certain placement strategy, the actual position of the placed components can be accurately predicted. Also, it is easy to notice that the prediction error of shift $X$ is lower than shift $Y$. This could be explained by the larger designed variance in $X$ direction (see Table 1) which can provide more information to train the prediction model. However, neither SVR-Linear or SVR-RBF model can estimate the shift angle well. It probably because some other key factors outside the 9 parameters are missed in our model. In terms of the two models, the SVR-RBF outperforms SVR-Linear considering the prediction errors: MAE and RMSE respectively. This indicates that the relationship between the 9 parameters and the component shifts in P&P process is not linear but non-linear. And the Gaussian-based SVR-RBF model is more suitable for the non-linear problems than SVR-Linear model.

Table 5. Prediction results of SVR-Linear and SVR-RBF models.

| Model | Response | MAE | RMSE |
| --- | --- | --- | --- |
| SVR-Linear | Shift $X$ (ratio) | 0.023 | 0.029 |
|  | Shift $Y$ (ratio) | 0.083 | 0.112 |
|  | Shift Angle (°) | 2.113 | 3.055 |
| SVR-RBF | Shift *X* (ratio) | 0.016 | 0.021 |
| | Shift *Y* (ratio) | 0.037 | 0.065 |
| | Shift Angle (°) | 1.798 | 2.748 |

## 5. Conclusions

In this research work, we have studied the component shifts in P&P process. A comprehensive experiment is designed and carried on a complete state-of-the-art SMT assembly line. 9 parameters of solder paste and placement are controlled or adjusted according to the theory of DOE. The SVR model is applied to predict the component shifts based on the two datasets: solder paste data (SPI) and component data (AOI). The feasibility of applying SVR methodology to predict the component shifts is evaluated. And the implementation of two kernel functions: SVR-Linear and SVR-RBF were used in the prediction model as well. Besides, mean absolute error (MAE) and root mean squared error (RMSE) are utilized to show the prediction accuracy and to test the performance of SVR-Linear model and SVR-RBF model. The results indicate that the component shifts in P&P process should not be ignored. And the SVR model is effective and attractive to estimate component shift X and shift Y with the high level of accuracy. Moreover, the SVR-RBF is superior to SVR-Linear in our case. However, the component shift angle is not addressed well in our case. To create a clearer and more effective prediction model of component shifts, especially for shift angle, we will still need to extend the scope of the involved factors to machinery, inspectional accuracy, environmental factors, etc. Thus, it will encompass more related factors to better estimate the component shifts in P&P process.

## Acknowledgments

The authors would like to thank the editors and the anonymous reviewers for your comments on this paper.